# Measurement of the Hubble constant using the Dark Energy Survey Year 6 Gold galaxy catalogue and the fourth Gravitational-Wave Transient Catalogue

Isaac McMahon,[1★] Danny Laghi,[1] Marcelle Soares-Santos,[1] Kendall Ackley,[2] Gergely Dálya,[3,4] Yavuz Gençel,[1] David Sánchez-Cid,[1] Felipe Andrade-Oliveira,[1] Sean MacBride,[1] Christian Chapman-Bird,[5] Rachel Gray[6] and Alexander Papadopoulos[6]

[1]Physik-Institut, University of Zürich, Winterthurerstrasse 190, CH-8057 Zürich, Switzerland
[2]Department of Physics, University of Warwick, Coventry CV4 7AL, UK
[3]L2IT, Laboratoire des 2 Infinis – Toulouse, Université de Toulouse, CNRS/IN2P3, UPS, F-31062 Toulouse Cedex 9, France
[4]HUN-REN–ELTE Extragalactic Astrophysics Research Group, 1117 Budapest, Hungary
[5]Institute for Gravitational Wave Astronomy & School of Physics and Astronomy, University of Birmingham, Edgbaston, Birmingham B15 2TT, UK
[6]SUPA, University of Glasgow, Glasgow G12 8QQ, UK

Accepted 2026 June 17. Received 2026 June 17; in original form 2026 February 5

## ABSTRACT

Gravitational wave (GW) standard sirens enable independent measurements of the Hubble constant $H_0$. In the absence of electromagnetic counterparts, the 'dark siren' method statistically correlates GW events with potential host galaxies. We present a measurement of $H_0$ using GWTC-4.0 (the fourth Gravitational-Wave Transient Catalogue) combined with the Dark Energy Survey Year 6 Gold photometric galaxy catalogue. Using the GWCOSMO pipeline, we jointly infer cosmological and GW population parameters. We analyse the impact of galaxy catalogue properties on the inference, identifying significant features in the galaxy redshift distribution which can introduce biases. By restricting the galaxy catalogue to $0.05 < z < 0.50$ to maintain consistency with a uniform in comoving volume galaxy distribution, we obtain a result of $H_0 = 71.7^{+22.7}_{-18.3}\,\mathrm{km\,s^{-1}\,Mpc^{-1}}$ from dark sirens and $H_0 = 73.1^{+11.9}_{-8.5}\,\mathrm{km\,s^{-1}\,Mpc^{-1}}$ when combined with the bright siren GW170817. This study demonstrates the adaptation of deep galaxy catalogues for GW cosmology, highlighting key challenges and methodologies essential for maximizing the potential of next-generation galaxy surveys.

**Key words:** gravitational waves – catalogues – cosmological parameters – cosmology: observations.

## 1 INTRODUCTION

The field of gravitational wave (GW) cosmology has rapidly developed alongside the growing number of GW candidate detections by the LIGO–Virgo–KAGRA (LVK) GW detector network (A. G. Abac et al. 2025). The first part of the fourth observing run (O4a) ended on 2024 January 16, and data products from O4a were released in 2025 August, bringing the number of total GW detections to 218 (The LIGO Scientific Collaboration, the Virgo Collaboration & the KAGRA Collaboration 2025a). GWs from compact binary coalescences (CBCs) act as standard sirens (D. E. Holz & S. A. Hughes 2005), which, when combined with electromagnetic observations of potential host galaxies, can be used to infer cosmological parameters, such as the Hubble constant $H_0$ (B. F. Schutz 1986; W. Del Pozzo 2012). This results in a measurement of $H_0$ which is independent of other methodologies. With sufficient precision, GWs may help to resolve the $\gtrsim 5\sigma$ tension between $H_0$ measurements from the early Universe (N. Aghanim et al. 2020) and the late Universe (A. G. Riess et al. 2022).

For a given GW detection, if a unique host galaxy is positively identified through astronomical searches, the redshift of the host galaxy and the luminosity distance of the GW can be used to measure $H_0$ through the luminosity distance–redshift relation (A. Nishizawa 2017). This is known as a 'bright siren', and only one such event, GW170817, has been confirmed to date (B. P. Abbott et al. 2017). In the absence of unique host galaxy identification, $H_0$ can still be measured by other methods, of which we consider two in this work.

One method measures $H_0$ from features in the GW event candidate mass spectrum, extracting information from the redshift between the source frame masses and detected masses, known as a 'spectral siren' analysis (D. F. Chernoff & L. S. Finn 1993; D. Marković 1993; S. R. Taylor, J. R. Gair & I. Mandel 2012; W. M. Farr et al. 2019; J. M. Ezquiaga & D. E. Holz 2021, 2022; S. Mastrogiovanni et al. 2021; Z. Q. You et al. 2021; G. Pierra, A. Colombo & S. Mastrogiovanni 2025). This method requires accurate modelling of the CBC population distribution and samples cosmological parameters alongside population parameters that describe the shape of the mass spectrum and event merger rate (S. Mukherjee 2022; G. Pierra et al. 2024; A. Agarwal et al. 2025).

★ E-mail: isaac.mcmahon@physik.uzh.ch

Another method statistically weights potential host galaxies for each GW candidate, leveraging redshift information from existing galaxy catalogues (C. L. MacLeod & C. J. Hogan 2008; J. R. Gair et al. 2023; N. Borghi et al. 2024). This is known as the 'galaxy catalogue' or 'dark siren' method. Recent developments in dark siren codes such as GWCOSMO (R. Gray et al. 2020, 2023; R. Gray, C. Messenger & J. Veitch 2022) and ICAROGW (S. Mastrogiovanni et al. 2023, 2024) allow joint inference of cosmological parameters and GW population parameters simultaneously, combining the spectral sirens method and the galaxy catalogue method into one. This combination is also referred to as the 'dark siren' method, and is the one which we use in this work. Additionally, other studies have used galaxy catalogues and dark sirens to measure cosmological parameters using cross-correlation (M. Oguri 2016; S. Mukherjee & B. D. Wandelt 2018; S. Bera et al. 2020; S. Mukherjee et al. 2024; J. A. Ferri et al. 2025; I. Santiago de Matos et al. 2025).

The galaxy catalogue and dark sirens method have previously been used with various galaxy catalogues, including previous data releases of the Dark Energy Survey (DES). The analyses using DES only considered a single event, assuming a fixed population model and merger rate parameters (M. Soares-Santos et al. 2019; A. Palmese et al. 2020) or only used DES for one event out of many (B. P. Abbott et al. 2021). Other analyses (M. Fishbach et al. 2019; A. Finke et al. 2021; R. Abbott et al. 2023b; The LIGO Scientific Collaboration, the Virgo Collaboration & the KAGRA Collaboration 2025b) use GLADE (Galaxy List for the Advanced Detector Era) or GLADE+ (G. Dálya et al. 2018, 2022), which is less than 50 per cent complete at $z > 0.05$ in the $K$ band, too shallow to provide significant information at typical GW distances. Other analyses with the Dark Energy Spectroscopic Instrument (DESI; W. Ballard et al. 2023), DESI Legacy Survey (A. Palmese et al. 2023), and DELVE (DECam Local Volume Exploration survey; C. R. Bom et al. 2024) consider only a small selection of available GW events and assume fixed mass distribution and merger rate parameters.

In this work, we adapt the DES Year 6 (Y6) Gold photometric galaxy catalogue (K. Bechtol et al. 2026) for the dark siren method and obtain an $H_0$ measurement with galaxies out to $z = 0.5$ (this choice of redshift range is discussed in Section 3.5). We prepare the galaxy catalogue for the cosmology analysis by performing object selection and classification (Sections 2.2 and 3.3). We measure the luminosity function of DES galaxies out to $z = 0.5$ with photometric redshift to quantify the completeness of our galaxy sample for dark sirens (Section 3.4). We describe our choices for computing the line-of-sight (LOS) redshift prior (Section 3.2). We present our computed LOS redshift priors and describe the features in the redshift distribution that we observe (Section 3.7). Finally, we show our measurement for $H_0$ and discuss potential sources of bias that arise from making different choices in the analysis (Section 4). If an assumed reference cosmology is required, we adopt $H_0 = 70\,\mathrm{km\,s^{-1}\,Mpc^{-1}}$ and $\Omega_\mathrm{m} = 0.3065$ (P. A. R. Ade et al. 2016) unless otherwise stated, as these are the defaults used in our chosen software (see Section 3.1).

## 2 DATA

### 2.1 Gravitational-wave events

The LVK network of observatories (A. G. Abac et al. 2025) consists of two Laser Interferometer Gravitational-Wave Observatory (LIGO; J. Aasi et al. 2015) detectors in the USA and the Virgo Observatory (F. Acernese et al. 2015) in Italy, as well as the KAGRA Observatory (T. Akutsu et al. 2021) in Japan. The fourth Gravitational-Wave Transient Catalogue (GWTC-4.0; A. G. Abac et al. 2025; The LIGO Scientific Collaboration, the Virgo Collaboration & the KAGRA Collaboration 2025a, c), consisting of approximately 7 months of data from 2023 May 24 to 2024 January 16, is the latest set of GW candidate data released by LVK. When combined with the previous GWTC-2.1 and GWTC-3.0 (R. Abbott et al. 2023a, 2024) data releases, they consist of all GW candidate data released by LVK to date. We use a candidate list which is consistent with the O4a LVK cosmology analysis (The LIGO Scientific Collaboration, the Virgo Collaboration & the KAGRA Collaboration 2025b), totalling 142 CBC GW candidates. 137 of these candidates are believed to have an origin in a binary black hole collision, and the remaining five candidates could have a neutron star component. The event GW170817, one of these five, is the only event with a confirmed electromagnetic counterpart (D. A. Coulter et al. 2017; M. Soares-Santos et al. 2017), and is thus treated separately in this analysis as a bright siren.

The GW information, which is used as an input for our analysis, comes in the form of parameter estimation posterior samples and HEALPIX skymaps (K. M. Gorski et al. 2005). The posterior samples encode information obtained only from the GW signal itself, including location, mass, spin, and other parameters that affect the gravitational waveform. The skymap is created from these samples and encodes posterior probability information both on the projected celestial sphere as well as luminosity distance in the radial direction (L. P. Singer & L. R. Price 2016). For all events, we use the publicly available data released on Zenodo (relevant links found in Section 5).

In addition to the real GW candidate data, we require simulated GW injections to model GW selection effects due to detector sensitivity (R. Essick et al. 2025). We use the injections with simulated detector sensitivity up to O4a, which were also released with GWTC-4.0.

### 2.2 Galaxy catalogue

The DES (The Dark Energy Survey Collaboration 2005; T. Abbott et al. 2016) was conducted using the Dark Energy Camera (DECam) at the 4-m Blanco telescope at Cerro Tololo Inter-American Observatory in Chile. The Y6 Gold data release (K. Bechtol et al. 2026) is curated for cosmology analysis, covering 4923 deg$^2$ in the southern sky and achieving a photometric uniformity of $<$2 mmag in the *griz* photometric filters over a total of 669 million objects. The 10$\sigma$ multi-epoch galaxy magnitude depth in $r$ band is 23.9. Included in the catalogue are photometric redshift (photo-$z$) measurements from DNF (Directional Neighbourhood Fitting; J. De Vicente, E. Sánchez & I. Sevilla-Noarbe 2016), with state-of-the-art average uncertainties $\sigma_z/(1+z) < 3$ per cent for $z < 1$. Using the $r$-band luminosity function (see Section 3.4) and the $r$-band apparent magnitude limit, we estimate that DES is approximately 100 per cent complete to $z \sim 0.5$ for an absolute magnitude limit $M_r < -17.5$ (M. Soares-Santos et al. 2019).

The data release includes three different models for photometric flux measurements: PSF (point spread function), bulge+disc fixed (BDF), and GAp (Gaussian aperture). In this analysis, we choose BDF as the representative photometry because the model performs better for galaxies. Specifically, we use the `BDF_MAG_*_CORRECTED` columns, which have been normalized to the global `MAG_APER_8` photometric calibration system and dereddened for Galactic extinction.

We apply the following recommended cuts based on quality flags for each source:

(i) Non-NaN redshift and photometry in the *griz* filters
(ii) `FLAGS_FOOTPRINT` == 1
(iii) `FLAGS_FOREGROUND` == 0
(iv) `FLAGS_GOLD` == 0

`FLAGS_FOOTPRINT` is a flag that equals 1 for all sources that have data from at least two exposures in each of the *griz* bands, defined by an NSIDE = 16 384 HEALPIX map of exposure areas. A cut of `FLAGS_FOOTPRINT` == 1 excludes 3.3 per cent of the full catalogue. `FLAGS_FOREGROUND` is a flag for sources that are impacted by the presence of nearby bright astrophysical foreground objects. The minimum area masked by the foreground mask is the area of one HEALPIX NSIDE = 4096 pixel, 0.86 arcmin$^2$. A cut of `FLAGS_FOREGROUND` == 0 excludes 9.7 per cent of the full catalogue. `FLAGS_GOLD` is a flag for photometric processing failures, objects with unphysical colours, noisy detections, and other issues. A cut of `FLAGS_GOLD` == 0 excludes 8.1 per cent of the full catalogue. In total, these quality cuts combine to exclude 19.1 per cent of the sources in the full catalogue.

Out of all 141 GW event candidates used in our analysis as dark sirens, 14 candidates (10 per cent) have more than 50 per cent probability to be in the DES footprint (see Fig. 1, upper panel). These candidates are well distributed among the redshifts to which DES is most sensitive. A total of ∼14 per cent of the probability of all 141 candidates falls within the DES footprint. Although most of the candidates fall outside of the DES survey footprint or redshift depth, we still include them in our analysis as they provide crucial cosmology information mainly through the shape of the GW mass distribution.

## 3 METHODOLOGY

### 3.1 Analysis framework

We jointly infer cosmology and CBC population parameters using a Bayesian framework as implemented in the GWCOSMO PYTHON package (R. Gray et al. 2022, 2023). Given a set of GW event data $\{d\}$ from a number $N_{\rm det}$ of detections, $\theta^{\rm det}$ are the parameters of the detected CBCs measured only from the GW data in the detector frame of reference. The joint posterior probability on CBC population parameters $\Lambda$ and cosmological parameters $\Lambda_{\rm c}$ given a set of GW events (individually labelled with $i$) can thus be written as

$$p(\Lambda, \Lambda_{\rm c}|\{d\}, N_{\rm det}) \propto \pi(\Lambda)\pi(\Lambda_{\rm c})\xi(\Lambda, \Lambda_{\rm c})^{-N_{\rm det}} \times \prod_{i=1}^{N_{\rm det}} \int {\rm d}\theta_i^{\rm det} \frac{p\left(\theta_i^{\rm det}|d_i\right)}{\pi(\theta_i^{\rm det})} \times \left|\frac{{\rm d}\theta_i^{\rm det}(\theta_i^{\rm src}, \Lambda_{\rm c})}{{\rm d}\theta_i^{\rm src}}\right|^{-1} p(\theta_i^{\rm src}|\Lambda), \quad (1)$$

where $\pi$ denotes a prior and the GW source-frame parameters $\theta^{\rm src}$ are given by a function of the detector-frame parameters and the cosmological parameters, $\theta^{\rm src} = \theta(\theta^{\rm det}, \Lambda_{\rm c})$. The term $\xi(\Lambda, \Lambda_{\rm c})$ is the expected fraction of GW events that are detected, required because the GW detectors will not be able to measure the entire true population of GW events due to some selection effects. It can be written as

$$\xi(\Lambda, \Lambda_{\rm c}) = \int {\rm d}\theta^{\rm det}\, P({\rm det}|\theta^{\rm det}) \times \left|\frac{{\rm d}\theta^{\rm det}(\theta^{\rm src}, \Lambda_{\rm c})}{{\rm d}\theta^{\rm src}}\right|^{-1} p(\theta^{\rm src}|\Lambda), \quad (2)$$

where $P({\rm det}|\theta^{\rm det})$ is the probability of detection given the measured GW parameters $\theta^{\rm det}$, dependent on the sensitivity of the detectors. The expected true underlying distribution of events $p(\theta^{\rm src}|\Lambda)$ depends on the chosen model for the population of CBC mergers and its parameters $\Lambda$.

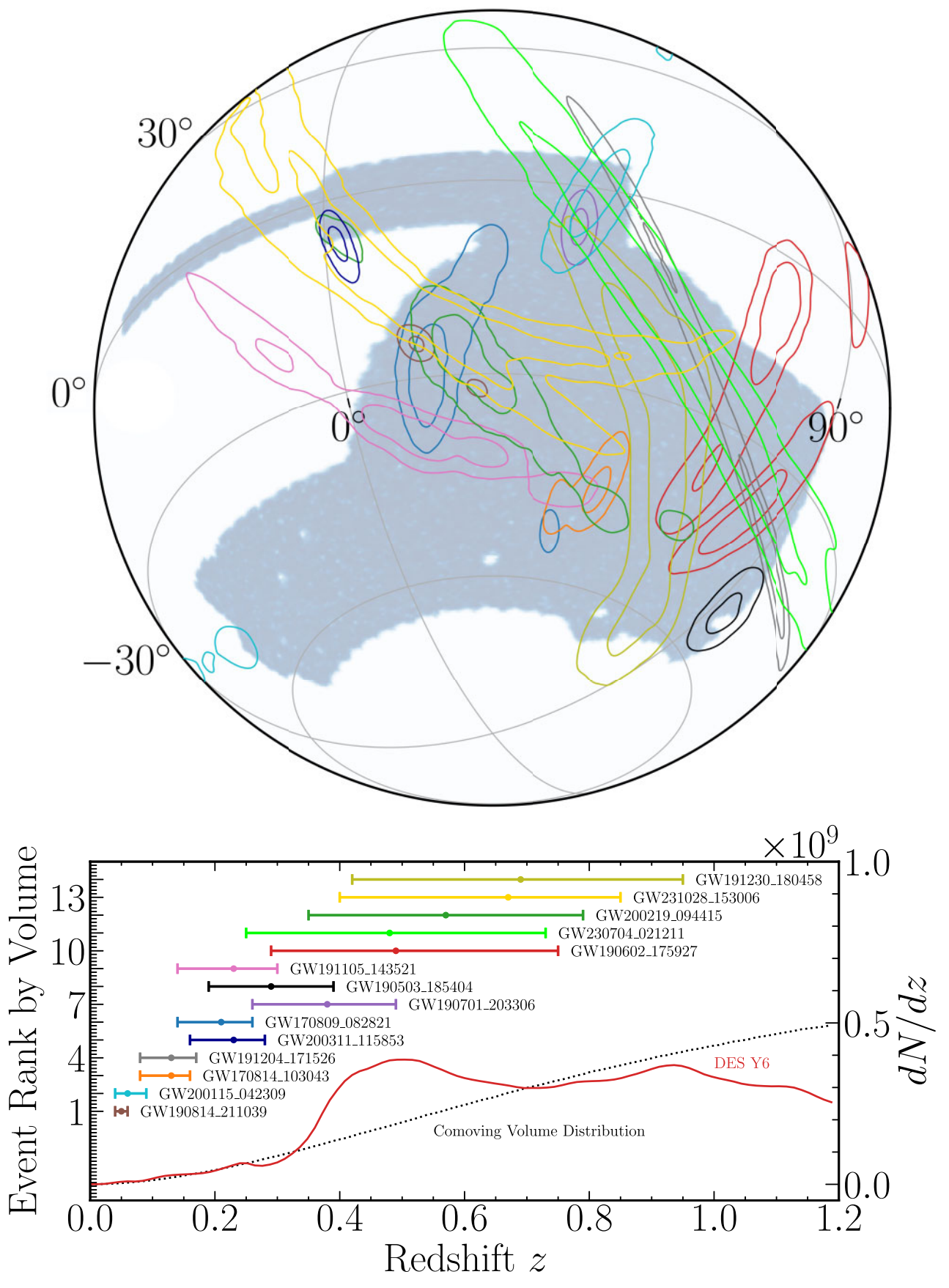

**Figure 1.** Top: GW event candidate 50 per cent and 90 per cent probability contours included in our analysis, which have more than 50 per cent probability to be within the DES Y6 Gold survey footprint (shown in blue). The axes are right ascension and declination in degrees. Bottom: The photometric redshift distribution of galaxies $dN/dz$ in DES Y6 Gold (red) compared to a uniform in comoving volume distribution (dotted black), scaled to have the same total number of galaxies. More discussion can be found in Section 3.7. On the same *x*-axis are shown redshift measurements (90 per cent confidence interval) for the events, converted from luminosity distance and ordered by increasing localization volume.

### 3.2 Line-of-sight redshift prior

In equation (1), the probability of a set of GW parameters $\theta_i^{\rm src}$ given $\Lambda$ depends on the distribution of potential host galaxies:

$$p(\theta_i^{\rm src}|\Lambda) \propto \frac{{\rm d}N^{\rm eff}_{\rm gal,cat}}{{\rm d}z\,{\rm d}\Omega} + \frac{{\rm d}N^{\rm eff}_{\rm gal,out}}{{\rm d}z\,{\rm d}\Omega}. \quad (3)$$

The first term ${\rm d}N^{\rm eff}_{\rm gal,cat}/{\rm d}z\,{\rm d}\Omega$ represents the spatial distribution of galaxies in a chosen galaxy catalogue across redshift $z$ and angular sky position $\Omega$. Each galaxy is represented with a Gaussian likelihood defined by its photometric redshift $z$ and the redshift error $\sigma_z$ at its sky location $\Omega$. While photometric redshift uncertainties are not truly Gaussian in general (A. Palmese et al. 2020, 2023; C. Turski et al. 2023; C. R. Bom et al. 2024), we use this approximation in this work.

The Gaussian likelihood of each galaxy is assigned a host probability weight $w_j$ as defined by

$$w_j\,(\epsilon, M_j) = 10^{-0.4\,\epsilon\,(M_j - M^*)}, \quad (4)$$

where $M^*$ is the reference absolute magnitude of the Schechter luminosity function, described in Section 3.4. The parameter $\epsilon$ is chosen to change how the galaxies are weighted with respect to each other. When $\epsilon = 0$, all galaxies are weighted equally as potential host galaxies for the GW event. When $\epsilon = 1$, this corresponds to a logarithmic weight in absolute magnitude, or equivalently, a linear weight in luminosity. The extent to which the probability of a galaxy being a host of a CBC merger may scale with its observables is unknown and is an active field of research (C. J. Neijssel et al. 2019; S. Adhikari et al. 2020; F. Santoliquido et al. 2021; F. S. Broekgaarden et al. 2022; L. Rauf et al. 2023; R. Srinivasan et al. 2023; A. Hanselman et al. 2025; G. Perna, S. Mastrogiovanni & A. Ricciardone 2025; Z. Li, R. Gray & I. S. Heng 2026). As host probabilities for stellar-origin CBCs are expected to scale with the stellar mass of the galaxy, and galaxy stellar mass roughly scales with luminosity (B. Mobasher et al. 2015), host probabilities are expected to also scale with luminosity. Most dark siren analyses to date consider only the cases $\epsilon = 0$ or $\epsilon = 1$, with $\epsilon = 1$ being the preferred fiducial analysis. In this work, we choose $\epsilon = 1$.

The second term $\mathrm{d}N^{\mathrm{eff}}_{\mathrm{gal,out}}/\mathrm{d}z\,\mathrm{d}\Omega$, known as the 'empty catalogue' case, represents galaxies that are assumed to exist but are not present in the catalogue due to selection effects. This term is modelled by assuming that all galaxies that are not detected by the catalogue follow an isotropic distribution, which is uniform in comoving volume. While this does not account for galaxy clustering, we expect that this model adequately describes the real average galaxy distribution over large scales.

In GWCOSMO, these terms are passed to the likelihood as a 'LOS redshift prior' which encodes the sum of the in-catalogue and out-of-catalogue terms as a probability distribution in redshift for every pixel in a HEALPIX map of chosen resolution (R. Gray et al. 2022). For each pixel, only the catalogued galaxies in that pixel contribute to the probability distribution. This LOS redshift prior is unique for each galaxy catalogue, and so is the focus of this work. We compute the LOS redshift prior for DES Y6 Gold using GWCOSMO and use it to obtain the posterior probability distributions on the population parameters $\Lambda$ and cosmological parameters $\Lambda_c$. While we do assume a cosmology in order to compute this LOS prior, the bulk of its dependence on $H_0$ materializes in a normalization constant which cancels out in the likelihood evaluation. There is a minor cosmology dependence in the catalogue completeness term, but we do not expect this to have a large impact our results (R. Gray et al. 2023).

The spacing of the redshift array used in the calculation of the redshift prior and the dark siren likelihood must be chosen to allow for sufficient resolution to capture information from every galaxy, considering its redshift error. The spacing between consecutive elements of the array must be smaller than the smallest redshift errors in the catalogue at that redshift. We adopt a redshift array that satisfies these conditions for the properties of the DES catalogue, with a linear spacing of $4.4 \times 10^{-4}$ for $0 < z \leq 0.5$, then logarithmic from $0.5 < z < 10$ to a maximum spacing of $8.9 \times 10^{-3}$.

### 3.3 Galaxy catalogue choices

Several choices must be made before computing the LOS redshift prior for DES Y6 Gold. We choose to use $r$-band photometry because the average photometric magnitude errors in $r$ are smaller than in the other three bands.

The catalogue provides two morphological classifiers to remove stars from our sample: the discrete-valued flag `EXT_MASH` [based on SOURCEEXTRACTOR (E. Bertin & S. Arnouts 1996) shape measurement outputs] and the continuous-valued `XGB_PRED` (made using the gradient boosted decision tree machine learning algorithm XGBOOST; T. Chen & C. Guestrin 2016). While `EXT_MASH` $== 4$ is the recommended fiducial classifier choice used in DES cosmological analyses, we use the object classification cut `EXT_XGB` $== 4$, defined as a cut at the value of `XGB_PRED`, which passes the same number of objects as `EXT_MASH` $== 4$. This is because XGBOOST classification performs better for objects with bright apparent magnitudes. Bright stars that are misclassified as galaxies will have large, spurious luminosities. This has an outsized impact on the dark siren redshift prior when luminosity weighting is used, causing large weights to be assigned to objects that are not galaxies.

While DES Y6 Gold has high completeness out to $z \sim 1$, we find that the photometric redshift distribution of galaxies in the catalogue does not closely follow our assumption of uniform in comoving volume in a region centred around $z \sim 0.5$ (see Fig. 1, bottom panel). This apparent overdensity arises from strong photo-$z$ bias in the range $0.15 < z < 0.35$ scattering into higher redshifts. This bias is caused by the lack of $u$-band measurements, the transition of the 4000 Å break from $g$ to $r$ at $z \sim 0.35$, and the lack of faint red galaxies at this redshift used for photo-$z$ estimation (C. Sánchez et al. 2014; J. De Vicente et al. 2016; M. Aguena et al. 2021). This bias is magnitude-dependent, reaching a bias of 0.1-0.2 for faint galaxies with $m_r > 20$, motivating an apparent magnitude cut $m_r < 20$ to mitigate it. We further discuss the features in the galaxy redshift distribution in Section 3.7.

Additionally, objects begin to saturate in the DECam sensors at apparent magnitudes of $m_r < 15$, meaning that objects which appear brighter than this are not catalogued in DES Y6 Gold. This saturation effect disproportionately affects galaxies at redshifts $z < 0.05$. For this reason, we also apply a redshift cut on the low end of the LOS redshift prior at $z_{\mathrm{min}} = 0.05$, excluding all in-catalogue contributions to the redshift prior below this redshift. For all redshifts outside of the interval defined by the redshift cuts, $z_{\mathrm{min}} < z < z_{\mathrm{max}}$, the redshift prior consists only of the out-of-catalogue term and follows exactly a uniform in comoving volume distribution. This has the added benefit of avoiding the need for bulk peculiar velocity corrections for the catalogue, as these corrections are only valid for redshifts $z < 0.05$ and are much smaller than typical photometric redshift errors (S. Mukherjee et al. 2021). No events in the DES footprint have the majority of their localization volume below $z < 0.05$, although two events have their localization volumes near the cut. As the comoving volume assumption is used in the cut region, the lost information does not bias the resulting measurement.

We choose a HEALPIX resolution of NSIDE = 128. For this resolution, pixels in the DES footprint have $\mathcal{O}(10^{2\text{-}3})$ galaxies between the chosen redshift cuts compared to $\mathcal{O}(10^{0\text{-}1})$ for GLADE+. Beyond this resolution, the improvement in the posterior from better resolution is marginal because the posterior is more limited by GW data precision than galaxy catalogue precision. We also adopt the same resolution for the coarse-resolution maps of the magni-tude threshold $m_{\rm thr}$ and effective galaxy number $N_{\rm eff}$, which are used for the calculation of the redshift prior, as our galaxy density allows us to be robust against small-number statistics even at such a resolution.

### 3.4 Luminosity function

The galaxy luminosity function, in the form of the Schechter function (P. Schechter 1976), is used in the calculation of the redshift prior to find the completeness fraction of the galaxy catalogue at a given redshift, which is used in the out-of-catalogue term of equation (3). The Schechter function in terms of absolute magnitude takes the form

$$\Phi(M)\,\mathrm{d}M = 0.4\ln(10)\,\phi^*\,10^{0.4\,(\alpha+1)(M^*-M)} \times \exp\left[-10^{0.4\,(M^*-M)}\right]\mathrm{d}M, \tag{5}$$

where $\phi^*$, $M^*$, and $\alpha$ are empirical characteristic parameters which describe the shape of the luminosity function. These parameters are passed as inputs to GWCOSMO to compute the LOS redshift prior, along with the maximum and minimum magnitudes, $M_{\rm max}$ and $M_{\rm min}$, which determine the interval on which the Schechter parametrization is trustworthy.

The galaxy luminosity function is known to evolve as a function of redshift (M. R. Blanton et al. 2003; J. Loveday et al. 2012). Some studies have also suggested that the luminosity function is better described by a double Schechter function (M. R. Blanton et al. 2005; D. Capozzi et al. 2017; J. F. Gschwend 2019). In GWCOSMO, the input Schechter parameters are assumed to describe a non-evolving single Schechter function. For this reason, we must use a single Schechter function and model evolution effects separately. To compute the redshift prior, we use the Schechter parameter values at the median redshift of our galaxy sample. We model the evolution effects to be linear in redshift and take the forms

$$\begin{aligned} M^*(z) &= M^*_0 + Q_M\,(z - z_0), \\ \alpha(z) &= \alpha_0 + Q_\alpha\,(z - z_0), \\ \phi^*(z) &= \phi^*_0\,10^{P_\phi\,(z-z_0)}, \end{aligned} \tag{6}$$

where $Q_M$, $Q_\alpha$, and $P_\phi$ are evolutionary parameters and $z_0$ is a zero-point redshift against which the parameters are corrected. There have been some studies that have measured Schechter function parameters for previous DES data releases (D. Capozzi et al. 2017; J. F. Gschwend 2019), but they only report parameters for the $i$ band. There have also been several studies in the last few decades which measure Schechter function parameters for the $r$ band (M. R. Blanton et al. 2003; A. D. Montero-Dorta & F. Prada 2009; D. T. Hill et al. 2010; J. Loveday et al. 2012), but all of these studies either ignore evolution effects, do not cover our desired redshift range, or report only the 'true' Schechter parameters at $z_0$ after correcting for evolution over the entire redshift range. For this reason, we empirically measure the Schechter parameters from DES Y6 Gold and use these to compute the redshift prior.

To measure the luminosity function, we adopt the $1/V_{\rm max}$ method (S. Eales 1993). We define the luminosity function $\Phi$ for a redshift range $z_1 < z < z_2$ to take the form

$$\int_{M_1}^{M_2} \Phi\,\mathrm{d}M = \sum_i^N \frac{P_i}{C_{s,i} V_{\mathrm{max},i}} \tag{7}$$

for an absolute magnitude bin with edges $M_1$ and $M_2$ which contains $N$ galaxies.

$P_i$ is the probability that the $i$th galaxy lies within the range $[z_1,\, z_2]$. Since only point-estimate photo-$z$s are available for DES and not full probability density functions (PDFs), we approximate each galaxy's PDF as a Gaussian function $\mathcal{N}$ defined by $z_i$ and $\sigma_{z,i}$:

$$P_i = \int_{z_1}^{z_2} \mathcal{N}(z,\, z_i,\, \sigma_{z,i})\,\mathrm{d}z. \tag{8}$$

$C_{s,i}$ is a galaxy completeness correction, with a value from 0 to 1, that corrects for the detection rate of galaxies in DES as a function of apparent magnitude and surface brightness (D. Capozzi et al. 2017; J. F. Gschwend 2019). This term was computed using the BALROG Y6 synthetic source injection catalogue (D. Anbajagane et al. 2025), a companion data set which characterizes object detection for DES Y6. Using synthetic BALROG galaxies, we interpolate the detection rate $C_s$ as a function of apparent magnitude and surface brightness.

The maximum accessible comoving volume for the $i$th galaxy $V_{\mathrm{max},i}$ is defined as

$$V_{\mathrm{max},i} = A_{\rm cat}\left[\frac{c}{H_0}\int_{z_{\mathrm{gal,min},i}}^{z_{\mathrm{gal,max},i}} \frac{\mathrm{d}z'}{\sqrt{\Omega_{\rm m}(1+z')^3 + (1-\Omega_{\rm m})}}\right]^3, \tag{9}$$

where $A_{\rm cat}$ is the solid angle subtended by the galaxy catalogue, $c$ is the speed of light, $H_0 = 100\,h$ is the Hubble constant, and $\Omega_{\rm m}$ is the cosmological matter density parameter. For measuring the luminosity function, it is customary to set $H_0 = 100\,\mathrm{km\,s^{-1}\,Mpc^{-1}}$ ($h = 1$), as the measured Schechter parameters can be easily converted to different $H_0$.

The limits of integration in equation (9) define the maximum redshift range of the $i$th galaxy, given the absolute magnitude of the galaxy and are defined as

$$z_{\mathrm{gal,min},i} = \max(z_1,\, z_{\mathrm{bright},i}), \quad z_{\mathrm{gal,max},i} = \min(z_2,\, z_{\mathrm{faint},i}), \tag{10}$$

where $z_{\mathrm{bright},i}$ and $z_{\mathrm{faint},i}$ are the redshifts for which a galaxy of absolute magnitude $M$ would have an apparent magnitude of the bright- and faint-end apparent magnitude limits of the galaxy catalogue, taken for DES Y6 Gold to be $m_{\mathrm{faint},r} = 23.9$ and $m_{\mathrm{bright},r} = 14$. The absolute magnitude $M$, apparent magnitude $m$, and $z$ of a given galaxy are related to each other by

$$M_r = m_r - 5\log_{10}\left[\frac{D_{\rm L}}{10\,\mathrm{pc}}\right] - K(z), \tag{11}$$

where $D_{\rm L}$ is the luminosity distance,

$$D_{\rm L} = \frac{c\,(1+z)}{H_0}\int_0^z \frac{\mathrm{d}z'}{\sqrt{\Omega_{\rm m}(1+z')^3 + (1-\Omega_{\rm m})}}, \tag{12}$$

and $K$ is the $K$ correction, described in Section 3.5. To estimate errors on the luminosity function, we subdivide the galaxy catalogue into 20 subregions using a *k-means* clustering algorithm (E. Suchyta et al. 2016; J. Kwan et al. 2017) to measure variations over the footprint of the survey, adjusting $A_{\rm cat}$ accordingly. We compute the mean and standard deviation value of $\Phi$ across all realizations and subregions for each magnitude bin and fit the Schechter function to these values.

We use this method to estimate the Schechter parameters of DES Y6 Gold and their evolution over the redshift range $0.05 < z < 0.5$, shown in Fig. 2 and Table 1. To determine the Schechter parameters to use to compute the redshift prior, we evaluate the Schechter parameters according to a linear evolution fit at the standard zero-point redshift $z_0 = 0.1$ and the median redshift of all galaxies in our DES sample, $z_0 = 0.24$, which is what we use to compute the LOS redshift prior. We define $M_{\rm min}$ and $M_{\rm max}$ by converting the upper and lower apparent magnitude limits of the survey to absolute magnitude using the mid-point of each redshift bin and the maximum $K$ correction allowed at that redshift (see Section 3.5). We also limit $M_{\rm max}$ to be no more than $M_{\rm max} \leq -18 + 5\log_{10}(h)$ to avoid problematic absolute magnitudes caused by the redshift bias of faint galaxies (see Section 3.3). These conservative limits cause our lowest redshift bin to have very poor constraints, so we exclude that bin from our evolution fit.

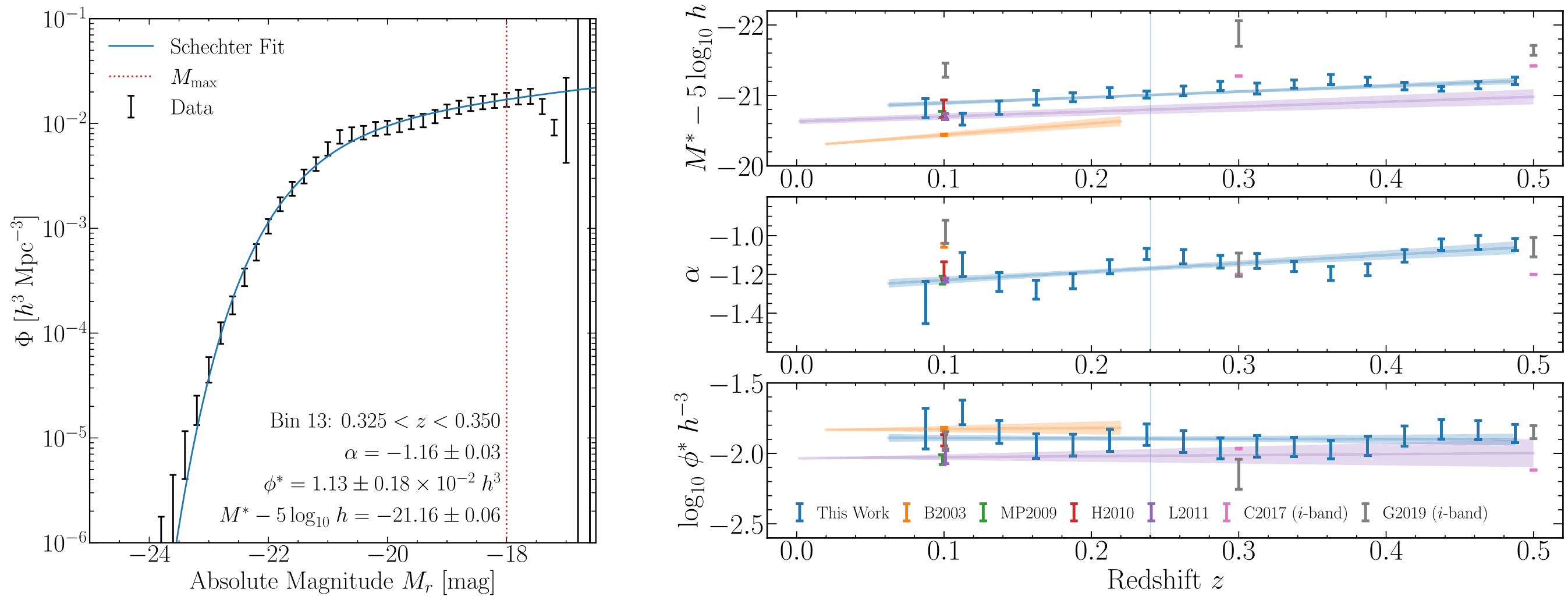

**Figure 2.** Left: Luminosity function with $1\sigma$ errors and corresponding Schechter function fit in the example redshift range $0.325 < z < 0.35$. $M_{\rm min}$ and $M_{\rm max}$ are shown with a dotted line. Right: $r$-band Schechter function parameters as a function of redshift and their linear evolution fits. Results from the literature are shown for comparison with evolution when included, typically measured at $z_0 = 0.1$, for B2003 (M. R. Blanton et al. 2003), MP2009 (A. D. Montero-Dorta & F. Prada 2009), H2010 (D. T. Hill et al. 2010), and L2011 (J. Loveday et al. 2012). Parameters measured with the $i$-band from C2017 (D. Capozzi et al. 2017) and G2019 (J. F. Gschwend 2019) are shown but are not directly comparable (C2017 also assumes $\alpha = -1.2$). The vertical blue line indicates the median redshift of the galaxy sample. The inferred Schechter parameters at the intersection of the blue lines are used to compute the redshift prior.

**Table 1.** Measured Schechter parameters for different zero-point redshifts.

| $z_0$ | $\phi^*$ ($10^{-2}\,h^3\,{\rm Mpc}^{-3}$) | $M^* - 5\log_{10} h$ (mag) | $\alpha$ | $M_{\rm min}$ (mag) | $M_{\rm max}$ (mag) |
|---|---|---|---|---|---|
| 0.10 | $1.28 \pm 0.09$ | $-20.89 \pm 0.03$ | $-1.23 \pm 0.02$ | $-22.26$ | $-18.00$ |
| 0.24 | $1.27 \pm 0.05$ | $-21.01 \pm 0.02$ | $-1.17 \pm 0.01$ | $-24.46$ | $-18.00$ |

We estimate the values of the linear evolution parameters to be $Q_M = -0.80 \pm 0.03$, $Q_\alpha = 0.43 \pm 0.01$, and $P_\phi = -0.03 \pm 0.02$. Our measurements are generally consistent with previous studies. Our results for $\alpha$ agree with the $i$-band results of J. F. Gschwend (2019), which is the most directly comparable study using a previous DES data release (D. Capozzi et al. 2017 assume a fixed value of $\alpha = -1.2$ for single-Schechter parameter results). We note that the evolution of $\alpha$ is poorly constrained and is often not modelled in such analyses. There have been some studies that suggest that $\alpha$ traces the redshift evolution of galaxy morphology (R. S. Ellis et al. 1996; O. Ilbert et al. 2005).

### 3.5 *K* corrections

The $K$ correction in equation (11) accounts for the shifting of the observed spectrum of a galaxy at redshift $z$, causing the portion of the spectrum which is detected by a given photometric filter to be different in the rest and observed frames (D. W. Hogg et al. 2002). In order to directly compare the absolute magnitudes of the galaxies in the rest frame, the magnitudes must be $K$-corrected. We adopt a polynomial fit of $K$ corrections in redshift and colour space (I. Chilingarian, A. L. Melchior & I. Zolotukhin 2010), allowing us to compute any $K$ correction for $r$ band given $z$ and $g - r$ colour. These $K$ corrections are only reliable for $z < 0.5$ and $-0.1 < g - r < 1.9$, imposed by the wavelength coverage of the chosen survey filters. For this reason, we impose these limits on our galaxy sample and our LOS prior calculation. The constraint imposed by $K$ correction calculations of $z < 0.5$ excludes ~80 per cent of the galaxies in DES, but this exclusion accounts for only ~24 per cent of the cumulative localization volume of all GW events up to GWTC-4.0 in the DES footprint. While the inclusion of higher redshift galaxies would further constrain $H_0$, the bulk of $H_0$ constraining power comes from $z < 0.5$.

The $K$ correction polynomials were originally fit to the Sloan Digital Sky Survey (SDSS) photometric filters, which slightly differ from DECam filters. Because of this, we convert the DECam magnitudes to SDSS magnitudes for all galaxies following a set of transformations defined by DES (T. M. C. Abbott et al. 2021).

### 3.6 Hardware acceleration of GWCOSMO

Previous large-scale analyses using GWCOSMO, such as those in The LIGO Scientific Collaboration, the Virgo Collaboration & the KAGRA Collaboration (2025b), have exclusively been performed via parallel processing over central processing unit (CPU) cores. However, as the number of included GW events grows, inference rapidly becomes computationally infeasible (with analyses of ∼200 events taking ≳1 week distributed over 32 CPU cores). This high computational cost is mainly due to the need to evaluate kernel density estimates (KDEs) over $\theta^{\rm src}$ for all sky-pixels associated with each GW event, which is necessary to compute the integral in equation (1). As $\theta^{\rm src}$ change for each new likelihood evaluation, these KDEs must be constructed on the fly. As open-source tools for parallel KDE construction are not available, GWCOSMO computes them sequentially, leading to further inefficiencies.

In this work, we employ a new version of GWCOSMO that both addresses this limitation and accelerates each likelihood evaluation with vectorization on GPU (graphics processing unit) hardware. The key structural change necessary to achieve this is to operate directly on a large three-dimensional array of posterior samples (with each axis referring to the event, pixel, or posterior sample index, respectively). As each pixel contains a different number of posterior samples, we pad the excess entries with NANS. This layout allows for high-level operations on posterior samples that can be easily expressed as array operations (e.g. parameter transformations, selection effect estimation, and population model evaluation) to be performed with the hardware-accelerated PYTORCH computational framework (A. Paszke et al. 2019), using NAN-aware operations where required. To maximize efficiency and limit memory consumption, we vectorize the KDE evaluation and subsequent integration of equation (1) with custom CUDA kernels, which we implement via the CUDA target of the NUMBA library (S. K. Lam, A. Pitrou & S. Seibert 2015). These kernels leverage low-level operations and shared memory usage to operate directly over the (non-padded) posterior samples, circumventing the instantiation of large intermediate arrays and minimizing wasted computations.

These improvements typically yield a speed-up of 1–2 orders of magnitude over a cluster of 32 CPU cores, depending on the number of GW events; this reduces analysis wall-times from days to hours, enabling GWCOSMO to scale to GW catalogues of hundreds (or even thousands) of events. Further details of the hardware acceleration of GWCOSMO can be found in A. Papadopoulos et al. (2026).

### 3.7 LOS prior validation

The resulting LOS redshift prior averaged over the DES footprint is shown in Fig. 3. For the regions $z < z_{\rm min}$ and $z > z_{\rm max}$, the prior reverts to the default uniform in comoving volume empty catalogue assumption. Although the galaxy catalogue, on average, is expected to align closely with the comoving volume assumption, we find that the prior constructed from the galaxy catalogue presents various features across the considered redshift range. We find that the average prior deviates from a comoving volume distribution by at most ∼21 per cent.

Some of these redshift features have been observed before and noted in previous dark siren studies. Earlier versions of the DES catalogue were used in single-event analyses of GW170814 (M. Soares-Santos et al. 2019) and GW190814 (A. Palmese et al. 2020), which found that a ‘galaxy wall’ overdensity structure exists in the DES footprint, spanning the region $35 < \alpha < 55$ and $-55 < \delta < -35$ and centred at $z \sim 0.06$. A similar peak was observed at $z \sim 0.12$. These features persist across multiple photometric redshift estimation algorithms and are also confirmed by several spectroscopic surveys of the region (S. A. Shectman et al. 1996; M. Colless et al. 2001; D. H. Jones et al. 2009), suggesting that these are real structures.

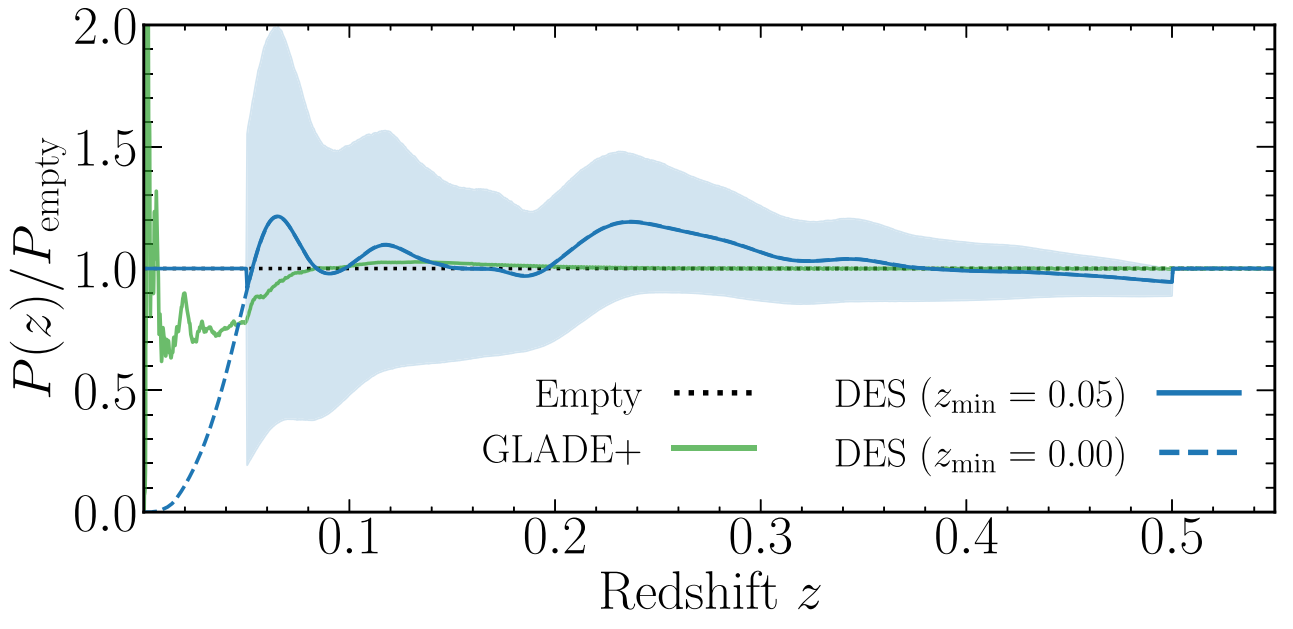

**Figure 3.** LOS redshift prior averaged over the footprint of the DES survey with upper and lower $1\sigma$ bounds. The $y$-axis is the ratio between the catalogue prior and the ‘empty’ catalogue $P_{\rm empty}$. The average LOS redshift prior for GLADE+ $K$-band is shown in green for comparison. The effect of applying a cut at $z_{\rm min}$ is shown with a dashed line at low redshift. If there is no cut, the prior dramatically decreases to zero due to the saturation selection effect.

**Table 2.** Posterior statistics for $H_0$ with $1\sigma$ (68.3 per cent) confidence intervals.

| Catalogue | $z_{\rm min}$ | $z_{\rm max}$ | $H_0$ (km s$^{-1}$ Mpc$^{-1}$) |
|---|---|---|---|
| DES | 0.05 | 0.50 | $71.7^{+22.7}_{-18.3}$ |
| DES | 0.00 | 0.50 | $78.3^{+23.2}_{-15.1}$ |
| Empty | – | – | $74.5^{+27.3}_{-20.7}$ |
| GLADE+ | – | – | $79.6^{+27.3}_{-18.5}$ |
| GW170817 | – | – | $78.2^{+26.0}_{-11.8}$ |
| DES+GW170817 | 0.05 | 0.50 | $73.1^{+11.9}_{-8.5}$ |

Large deviations from the assumption cause step features at the $z_{\rm max}$ boundary, where the in-catalogue term sharply transitions to the out-of-catalogue term. While this step feature may be an indicator of biases in the construction of the catalogue, which may affect the inference of $H_0$, the presence of the step feature itself does not add additional bias if it is within acceptable limits. This is because the likelihood calculation of the posterior is not sensitive to discontinuities in the redshift slope.

## 4 RESULTS

Using our calculated LOS prior, we run GWCOSMO to obtain joint posterior sample distributions for $H_0$ and GW population parameters, using the GW event catalogue and injections outlined in Section 2.1. For our mass population model, we use the FULLPOP-4.0 (The LIGO Scientific Collaboration, the Virgo Collaboration & the KAGRA Collaboration 2025d), which infers parameters of both neutron star and black hole populations at once. We assume Madau–Dickinson merger rate evolution (P. Madau & M. Dickinson 2014) and unmodified general relativity. We used the NESSAI nested sampler with normalizing flows (M. J. Williams, J. Veitch & C. Messenger 2021, 2023) with 1500 live points and default settings. We use identical priors for all parameters as those used in The LIGO Scientific Collaboration, the Virgo Collaboration & the KAGRA Collaboration (2025b). We also conduct spectral siren and GLADE+ runs to recover previous results. We report the results for $H_0$ in Table 2. More information about the mass model and results for all sampled parameters, including mass model parameters, can be found in Appendix A.

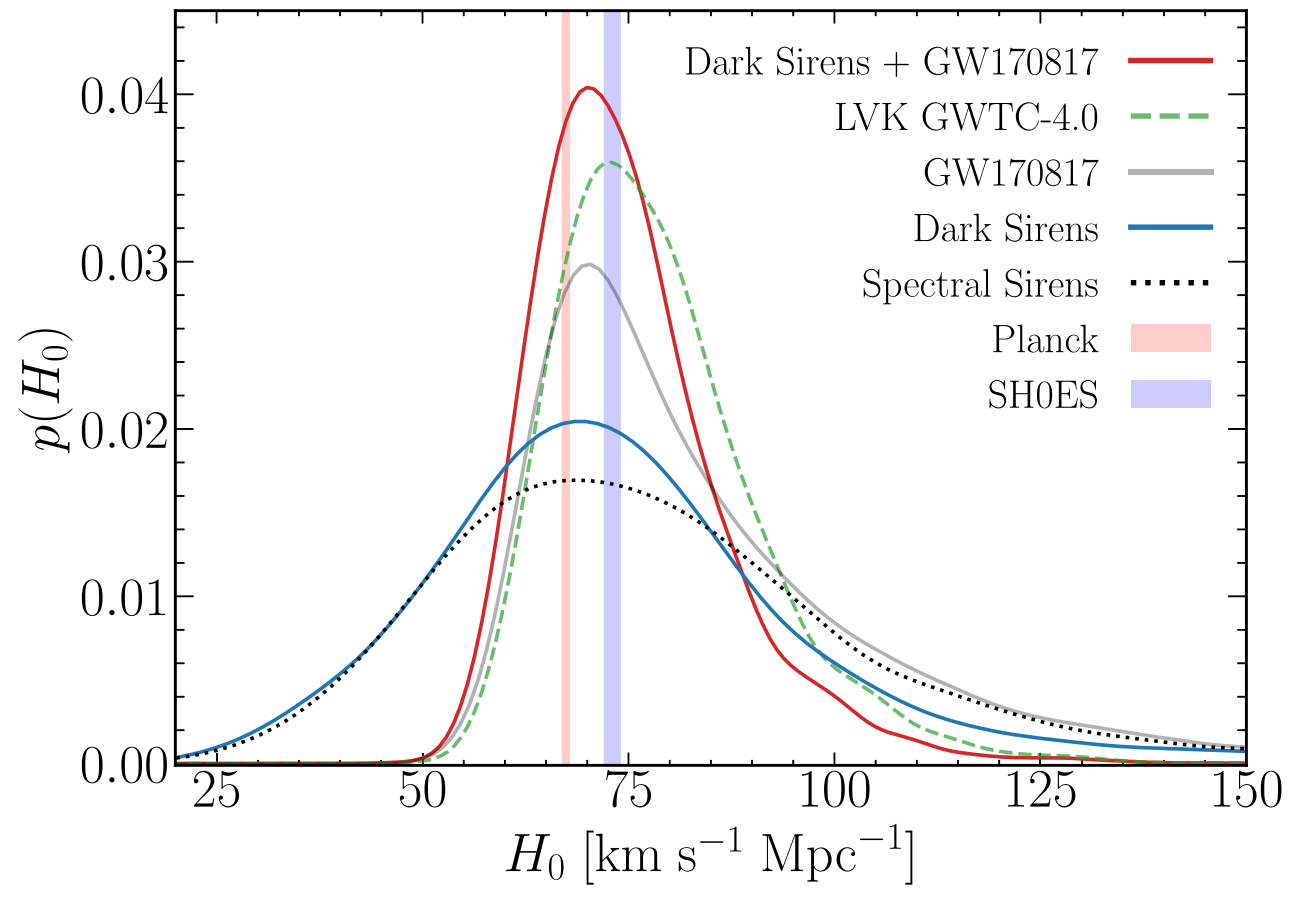

**Figure 4.** The final standard siren result of this work (red), made from the dark siren posterior with $0.05 < z < 0.5$ (blue) combined with the bright siren posterior for GW170817 (grey). The equivalent final combined result from LVK O4a for GLADE+ is shown in green (The LIGO Scientific Collaboration, the Virgo Collaboration & the KAGRA Collaboration 2025b), and the 'empty' catalogue spectral siren posterior is shown with a dotted line. $1\sigma$ regions for *Planck* (N. Aghanim et al. 2020) and SH0ES (A. G. Riess et al. 2022) are also shown.

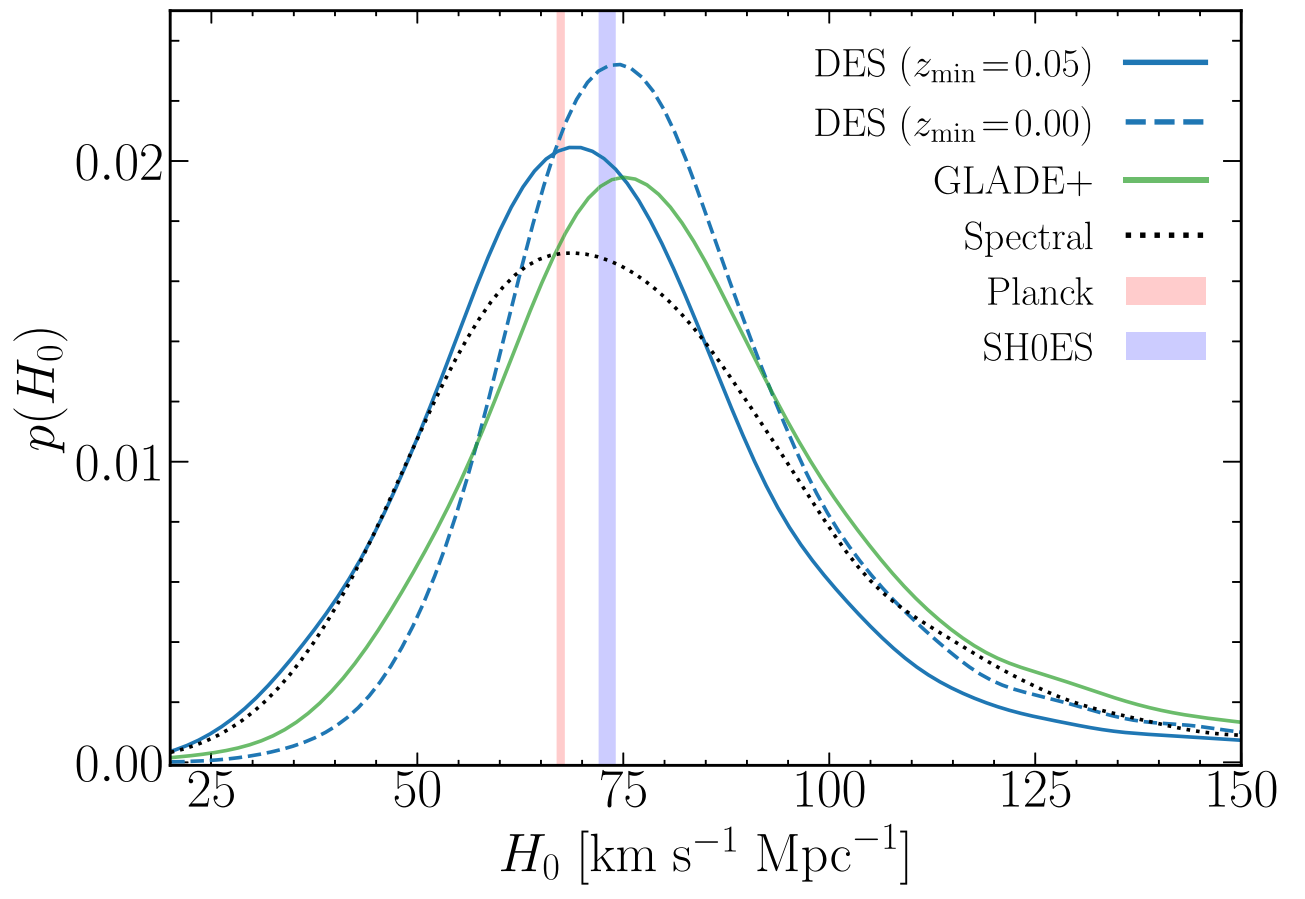

**Figure 5.** Posterior distributions with and without the use of $z_{\rm min}$. The dark siren posterior for GLADE+ is shown for comparison. $1\sigma$ regions for *Planck* (N. Aghanim et al. 2020) and SH0ES (A. G. Riess et al. 2022) are also shown.

Fig. 4 shows our final estimate for dark sirens and standard sirens. For the dark sirens only (with $0.05 < z < 0.50$), we obtain a measurement of $H_0 = 71.7^{+22.7}_{-18.3}$, representing a $\sim$11 per cent reduction in $1\sigma$ (68.3 per cent) confidence interval width compared to O4a dark sirens with GLADE+. We combine our results with the bright siren GW170817 posterior publicly released with the LVK O4a cosmology results and obtain a combined fiducial standard siren measurement of $H_0 = 73.1^{+11.9}_{-8.5}$, representing a similar $\sim$10 per cent improvement in precision over the O4a standard siren result.

The bulk of the constraining power comes from the coupling of cosmology to features in the mass spectrum of all the events, meaning that all the results from the galaxy catalogues currently provide only a small improvement to the empty catalogue spectral sirens case. This is because the galaxy catalogues are either too narrow or too shallow to provide much galaxy distribution information for most of the GW events in the sample. As galaxy catalogues improve, so will the cosmology constraints from dark sirens.

We show the effect of applying $z_{\rm min}$ to avoid selection effects from saturation in Fig. 5. In general, the deweighting of the prior at $z < 0.05$ causes the corresponding values of $H_0$ to be similarly deweighted, artificially pushing the measured peak of the $H_0$ distribution to higher redshifts for $z_{\rm min} = 0$.

As a systematic test, we also varied our choice of $M_{\rm max}$ from $-16$ to $-19$ to test the sensitivity of the $H_0$ result to the exclusion of these faint galaxies. We find that the median value of $H_0$ for each case does not shift, and the resulting constraint varies by no more than 5 per cent.

## 5 CONCLUSIONS

In this work, we demonstrate the use of large galaxy catalogues for joint dark siren analysis with cosmology and population parameters by making relevant object cuts and measuring the luminosity function. Analysing 141 GW events up to GWTC-4.0 as dark sirens with the DES Y6 Gold galaxy catalogue, we obtain a result of $H_0 = 71.7^{+22.7}_{-18.3}$ km s$^{-1}$ Mpc$^{-1}$ and $H_0 = 73.1^{+11.9}_{-8.5}$ km s$^{-1}$ Mpc$^{-1}$ when combined with the bright siren GW170817, using the FULLPOP-4.0 mass model and assuming host galaxy weighting by luminosity. This represents a $\sim$11 per cent improvement over the LVK cosmology dark siren analysis with GWTC-4.0 and the GLADE+ galaxy catalogue and a similar $\sim$10 per cent improvement when combined with GW170817. We show the importance of catalogue redshift depth by obtaining a competitive result with a galaxy catalogue that covers $\sim$12 per cent of the full sky and has a galaxy density of the order $\mathcal{O}(10^{2-3})$ times greater than GLADE+ .

We find that, because of limitations imposed by calculating $K$ corrections with the observed survey filters, we must exclude galaxies with redshift $z > 0.5$. Furthermore, we further impose a galaxy cut at $z > 0.05$ due to CCD saturation effects at low redshift. We show the effects of imposing these cuts on the resulting $H_0$ posterior. Additionally, DES only covers $\sim$12 per cent of the sky, limiting the overlap of GW events with the galaxy catalogue. We use a global Schechter function parametrization that is corrected for evolution based on the median redshift of galaxies in the given sample instead of smoothly modelling the Schechter evolution inside the LOS prior calculation. The evolution of the luminosity function with redshift is a lower order effect that none the less has important implications for the LOS prior (C. Turski et al. 2026).

Multiple unprecedented wide-field galaxy surveys are scheduled to be released in the next few years, such as DESI Data Release 2 (M. Abdul Karim et al. 2025), LSST (Large Synoptic Survey Telescope; Z. Ivezić et al. 2019), *Euclid* (L. Amendola et al. 2018), and 4MOST (4-metre Multi-Object Spectrograph Telescope; A. Verdier et al. 2026). The amount of galaxy data that is useful for dark sirens in both area and depth will soon grow immensely. The last two data releases for the fourth LVK observing run (O4) are expected to be released before the end of 2026 (N. Arnaud 2025). Plans for the fifth LVK observing run (O5) are not yet finalized but are expected to begin in late 2027 with major upgrades to all detectors, yielding even more GW detections, possibly including more bright siren candidates. For these reasons, this work and future improvements will be instrumental in preparing for the era of precision GW cosmology with these future releases.

## ACKNOWLEDGEMENTS

We thank Benoit Revenu for advice on analysis input parameters and the use of GWCOSMO. We thank the LVK cosmology working group for the productive discussions throughout the development of this work. We thank Archisman Ghosh and Maciej Bilicki for their contributions to the LVK internal review. We also thank the anonymous referee for the insightful review of our manuscript.

Funding for this work was provided by the University of Zurich (UZH) and the Swiss National Science Foundation (SNSF) under grant number 10002981. DL was supported by the UZH Postdoc Grant, grant number K-72341-01-01. CC-B acknowledges support from the United Kingdom Space Agency (UKSA) grant UKRI971. AP was supported by the United Kingdom Research and Innovation (UKRI) Science and Technology Facilities Council (STFC) studentship 323353–01.

This work made use of infrastructure services provided by the UZH Science IT team (www.s3it.uzh.ch). This material is based upon work supported by NSF's LIGO Laboratory, which is a major facility fully funded by the National Science Foundation. This project used public archival data from the Dark Energy Survey (DES). Funding for the DES Projects has been provided by the DOE and NSF (USA), MEC/MICINN/MINECO (Spain), STFC (UK), HEFCE (UK), NCSA (UIUC), KICP (U. Chicago), CCAPP (Ohio State), MIFPA (Texas A&M), CNPq, FAPERJ, FINEP (Brazil), DFG (Germany), and the DES Collaborating Institutions. Based in part on observations at Cerro Tololo Inter-American Observatory, NOIRLAB, which is operated by the Association of Universities for Research in Astronomy (AURA) under a cooperative agreement with the NSF.

## DATA AVAILABILITY

The DES Y6 Gold galaxy catalogue can be accessed at NCSA,[1] CosmoHub,[2] or NOIRLab.[3] GLADE+ can be accessed at its dedicated website.[4] GW parameter estimation samples can be downloaded on Zenodo for GWTC-4.0,[5] GWTC-3.0,[6] and GWTC-2.1,[7] as well as the injections used to model the GW selection effect[8] and cosmology posterior samples and distributions from the LVK O4a cosmology analysis.[9] We make use of the GWCOSMO PYTHON package.[10] All other code or data produced for this work can be shared upon reasonable request to the corresponding author.

## REFERENCES

Aasi J. et al., 2015, Class. Quantum Gravity, 32, 074001
Abac A. G. et al., 2025, ApJ, 995, L18
Abbott T. et al., 2016, MNRAS, 460, 1270

[1]https://des.ncsa.illinois.edu/releases/y6a2/Y6gold
[2]https://cosmohub.pic.es/catalogs/333
[3]https://datalab.noirlab.edu/
[4]https://glade.elte.hu/
[5]https://zenodo.org/records/17014085
[6]https://zenodo.org/records/8177023
[7]https://zenodo.org/records/6513631
[8]https://zenodo.org/records/16740128
[9]https://zenodo.org/records/16919645
[10]https://git.ligo.org/lscsoft/gwcosmo

Abbott B. P. et al., 2017, Nature, 551, 85
Abbott B. P. et al., 2021, ApJ, 909, 218
Abbott T. M. C. et al., 2021, ApJS, 255, 20
Abbott R. et al., 2023a, Phys. Rev. X, 13, 041039
Abbott R. et al., 2023b, ApJ, 949, 76
Abbott R. et al., 2024, Phys. Rev. D, 109, 022001
Abdul Karim M. et al., 2025, Phys. Rev. D, 112, 083515
Acernese F. et al., 2015, Class. Quantum Gravity, 32, 024001
Ade P. A. R. et al., 2016, A&A, 594, A13
Adhikari S., Fishbach M., Holz D. E., Wechsler R. H., Fang Z., 2020, ApJ, 905, 21
Agarwal A. et al., 2025, ApJ, 987, 47
Aghanim N. et al., 2020, A&A, 641, A6
Aguena M. et al., 2021, MNRAS, 502, 4435
Akutsu T. et al., 2021, Prog. Theor. Exp. Phys., 2021, 05A101
Amendola L. et al., 2018, Living Rev. Relativ., 21, 2
Anbajagane D. et al., 2025, Open J. Astrophys., 8, 65
Arnaud N., 2025, LVK Update for OpenMMA Call. https://dcc.ligo.org/LIGO-G2500315/public ( accessed 2026 February 2)
Ballard W. et al., 2023, Res. Notes AAS, 7, 250
Bechtol K. et al., 2026, ApJS, 282, 62
Bera S., Rana D., More S., Bose S., 2020, ApJ, 902, 79
Bertin E., Arnouts S., 1996, A&AS, 117, 393
Blanton M. R. et al., 2003, ApJ, 592, 819
Blanton M. R., Lupton R. H., Schlegel D. J., Strauss M. A., Brinkmann J., Fukugita M., Loveday J., 2005, ApJ, 631, 208
Bom C. R., Alfradique V., Palmese A., Teixeira G., Santana-Silva L., Santos A., Darc P., 2024, MNRAS, 535, 961
Borghi N., Mancarella M., Moresco M., Tagliazucchi M., Iacovelli F., Cimatti A., Maggiore M., 2024, ApJ, 964, 191
Broekgaarden F. S. et al., 2022, MNRAS, 516, 5737
Capozzi D. et al., 2017, preprint (arXiv:1707.09066)
Chen T., Guestrin C., 2016, in KDD '16: Proceedings of the 22nd ACM SIGKDD International Conference on Knowledge Discovery and Data Mining. Association for Computing Machinery (ACM), New York, p. 785
Chernoff D. F., Finn L. S., 1993, ApJ, 411, L5
Chilingarian I., Melchior A. L., Zolotukhin I., 2010, MNRAS, 405, 1409
Colless M. et al., 2001, MNRAS, 328, 1039
Coulter D. A. et al., 2017, Science, 358, 1556
Dálya G. et al., 2018, MNRAS, 479, 2374
Dálya G. et al., 2022, MNRAS, 514, 1403
Del Pozzo W., 2012, Phys. Rev. D, 86, 043011
De Vicente J., Sánchez E., Sevilla-Noarbe I., 2016, MNRAS, 459, 3078
Eales S., 1993, ApJ, 404, 51
Ellis R. S., Colless M., Broadhurst T., Heyl J., Glazebrook K., 1996, MNRAS, 280, 235
Essick R. et al., 2025, Phys. Rev. D, 112 102001
Ezquiaga J. M., Holz D. E., 2021, ApJ, 909, L23
Ezquiaga J. M., Holz D. E., 2022, Phys. Rev. Lett., 129, 061102
Farah A., Fishbach M., Essick R., Holz D. E., Galaudage S., 2022, ApJ, 931, 108
Farr W. M., Fishbach M., Ye J., Holz D. E., 2019, ApJ, 883, L42
Ferri J. A., Tashiro I. L., Abramo L., Matos I., Quartin M., Sturani R., 2025, J. Cosmol. Astropart. Phys., 04, 008
Finke A., Foffa S., Iacovelli F., Maggiore M., Mancarella M., 2021, J. Cosmol. Astropart. Phys., 08, 026
Fishbach M. et al., 2019, ApJ, 871, L13
Fishbach M., Essick R., Holz D. E., 2020, ApJ, 899, L8
Gair J. R. et al., 2023, AJ, 166, 22
Gorski K. M., Hivon E., Banday A. J., Wandelt B. D., Hansen F. K., Reinecke M., Bartelmann M., 2005, ApJ, 622, 759
Gray R. et al., 2020, Phys. Rev. D, 101, 122001
Gray R., Messenger C., Veitch J., 2022, MNRAS, 512, 1127
Gray R. et al., 2023, J. Cosmol. Astropart. Phys., 12, 023
Gschwend J. F., 2019, PhD thesis, Observatório Nacional, Rio de Janeiro, Brazil
Hanselman A., Vijaykumar A., Fishbach M., Holz D. E., 2025, ApJ, 979, 9

Hill D. T., Driver S. P., Cameron E., Cross N., Liske J., Robotham A., 2010, MNRAS, 404, 1215
Hogg D. W., Baldry I. K., Blanton M. R., Eisenstein D. J., 2002, preprint (arXiv:astro-ph/0210394)
Holz D. E., Hughes S. A., 2005, ApJ, 629, 15
Ilbert O. et al., 2005, A&A, 439, 863
Ivezić Z. et al., 2019, ApJ, 873, 111
Jones D. H. et al., 2009, MNRAS, 399, 683
Kwan J. et al., 2017, MNRAS, 464, 4045
Lam S. K., Pitrou A., Seibert S., 2015, in LLVM '15: Proceedings of the Second Workshop on the LLVM Compiler Infrastructure in HPC. Association for Computing Machinery (ACM), New York, p. 1
Li Z., Gray R., Heng I. S., 2026, ApJ, 1003, 176
Loveday J. et al., 2012, MNRAS, 420, 1239
MacLeod C. L., Hogan C. J., 2008, Phys. Rev. D, 77, 043512
Madau P., Dickinson M., 2014, ARA&A, 52, 415
Mali U., Essick R., 2025, ApJ, 980, 85
Marković D., 1993, Phys. Rev. D, 48, 4738
Mastrogiovanni S. et al., 2021, Phys. Rev. D, 104, 062009
Mastrogiovanni S. et al., 2023, Phys. Rev. D, 108, 042002
Mastrogiovanni S. et al., 2024, A&A, 682, A167
Mobasher B. et al., 2015, ApJ, 808, 101
Montero-Dorta A. D., Prada F., 2009, MNRAS, 399, 1106
Mukherjee S., 2022, MNRAS, 515, 5495
Mukherjee S., Wandelt B. D., 2018, preprint (arXiv:1808.06615)
Mukherjee S., Lavaux G., Bouchet F. R., Jasche J., Wandelt B. D., Nissanke S., Leclercq F., Hotokezaka K., 2021, A&A, 646, A65
Mukherjee S., Krolewski A., Wandelt B. D., Silk J., 2024, ApJ, 975, 189
Neijssel C. J. et al., 2019, MNRAS, 490, 3740
Nishizawa A., 2017, Phys. Rev. D, 96, 101303
Oguri M., 2016, Phys. Rev. D, 93, 083511
Palmese A. et al., 2020, ApJ, 900, L33
Palmese A., Bom C. R., Mucesh S., Hartley W. G., 2023, ApJ, 943, 56
Papadopoulos A., Chapman-Bird C. E. A., Gray R., Messenger C., Bertheas T., 2026, preprint (arXiv:2605.23538)
Paszke A. et al., 2019, preprint (arXiv:1912.01703)
Perna G., Mastrogiovanni S., Ricciardone A., 2025, A&A, 698, A128
Pierra G., Mastrogiovanni S., Perriès S., Mapelli M., 2024, Phys. Rev. D, 109, 083504
Pierra G., Colombo A., Mastrogiovanni S., 2025, preprint (arXiv:2511.11795)
Rauf L., Howlett C., Davis T. M., Lagos C. D. P., 2023, MNRAS, 523, 5719
Riess A. G. et al., 2022, ApJ, 934, L7
Sánchez C. et al., 2014, MNRAS, 445, 1482
Santiago de Matos I., Dalang C., Baker T., Abramo R., Ferri J., Quartin M., 2025, preprint (arXiv:2512.15380)
Santoliquido F., Mapelli M., Giacobbo N., Bouffanais Y., Artale M. C., 2021, MNRAS, 502, 4877
Schechter P., 1976, ApJ, 203, 297
Schutz B. F., 1986, Nature, 323, 310
Shectman S. A., Landy S. D., Oemler A., Tucker D. L., Lin H., Kirshner R. P., Schechter P. L., 1996, ApJ, 470, 172
Singer L. P., Price L. R., 2016, Phys. Rev. D, 93, 024013
Soares-Santos M. et al., 2017, ApJ, 848, L16
Soares-Santos M. et al., 2019, ApJ, 876, L7
Srinivasan R., Lamberts A., Bizouard M. A., Bruel T., Mastrogiovanni S., 2023, MNRAS, 524, 60
Suchyta E. et al., 2016, MNRAS, 457, 786
Taylor S. R., Gair J. R., Mandel I., 2012, Phys. Rev. D, 85, 023535
The Dark Energy Survey Collaboration, 2005, preprint (arXiv:astro-ph/0510346)
The LIGO Scientific Collaboration, the Virgo Collaboration, the KAGRA Collaboration, 2025a, preprint (arXiv:2508.18082)
The LIGO Scientific Collaboration, the Virgo Collaboration, the KAGRA Collaboration, 2025b, preprint (arXiv:2509.04348)
The LIGO Scientific Collaboration, the Virgo Collaboration, the KAGRA Collaboration, 2025c, preprint (arXiv:2508.18081)
The LIGO Scientific Collaboration, the Virgo Collaboration, the KAGRA Collaboration, 2025d, preprint (arXiv:2508.18083)
Turski C., Bilicki M., Dálya G., Gray R., Ghosh A., 2023, MNRAS, 526, 6224
Turski C., Brozzetti M. L., Dálya G., Punturo M., Ghosh A., 2026, MNRAS, 546, stag218
Verdier A. et al., 2026, MNRAS, 545, 3
Williams M. J., Veitch J., Messenger C., 2021, Phys. Rev. D, 103, 103006
Williams M. J., Veitch J., Messenger C., 2023, Mach. Learn.: Sci. Technol., 4, 035011
You Z. Q., Zhu X. J., Ashton G., Thrane E., Zhu Z. H., 2021, ApJ, 908, 215

## APPENDIX A: ADDITIONAL POSTERIOR PARAMETERS

We adopt the FULLPOP-4.0 model for the distribution of primary CBC masses. This is a phenomenological model that combines theoretical expectations with previous GW observations to parametrize the distributions of both black holes and neutron stars in CBCs (M. Fishbach, R. Essick & D. E. Holz 2020; A. Farah et al. 2022; U. Mali & R. Essick 2025). The model features two separate power-law components for neutron stars and black holes, two Gaussian peaks within the black hole region, and a dip at the apparent mass gap between neutron stars and black holes. In total, there are 19 parameters in the mass model. More details about the definitions, priors, and construction of the parameters can be found in The LIGO Scientific Collaboration, the Virgo Collaboration & the KAGRA Collaboration (2025b).

The shape of the mass model with the parameters sampled in our fiducial result can be seen in Fig. A1. The measured values for all sampled parameters are found in Table A1. Almost all of the parameters change very little with different catalogues and analysis choices, with the exception of the parameters which describe the position of the limits and peaks of the mass spectrum ($m_{\rm max}$, $\mu_{\rm g}^{\rm low}$, and $\mu_{\rm g}^{\rm high}$). This is because these are the parameters that have the highest correlation with $H_0$. Otherwise, we find a high degree of consistency between our results and previous studies.

**Table A1.** Full posterior parameter results with 1$\sigma$ (68.3 per cent) confidence intervals.

| Parameter | Unit | DES ($z_{\rm min}=0.05$) | DES ($z_{\rm min}=0.00$) | GLADE+ | Spectral |
|---|---|---|---|---|---|
| $H_0$ | km s$^{-1}$ Mpc$^{-1}$ | $71.7^{+22.7}_{-18.3}$ | $78.3^{+23.2}_{-15.1}$ | $79.6^{+27.3}_{-18.5}$ | $74.5^{+27.3}_{-20.7}$ |
| $\gamma$ | | $3.35^{+0.76}_{-0.64}$ | $3.28^{+0.72}_{-0.61}$ | $3.27^{+0.72}_{-0.61}$ | $3.36^{+0.75}_{-0.63}$ |
| $z_{\rm p}$ | | $2.50^{+1.00}_{-0.95}$ | $2.52^{+0.98}_{-0.97}$ | $2.53^{+0.97}_{-0.96}$ | $2.46^{+1.00}_{-0.95}$ |
| $\kappa$ | | $2.8^{+2.0}_{-1.9}$ | $2.8^{+2.1}_{-1.9}$ | $2.9^{+2.0}_{-2.0}$ | $2.8^{+2.1}_{-1.9}$ |
| $\alpha_1$ | | $2.92^{+1.18}_{-0.93}$ | $2.95^{+1.20}_{-0.94}$ | $3.03^{+1.20}_{-0.94}$ | $2.95^{+1.22}_{-0.94}$ |
| $\alpha_2$ | | $1.65^{+0.33}_{-0.42}$ | $1.67^{+0.33}_{-0.42}$ | $1.64^{+0.35}_{-0.46}$ | $1.65^{+0.34}_{-0.44}$ |
| $\beta_1$ | | $1.11^{+0.59}_{-0.49}$ | $1.16^{+0.59}_{-0.50}$ | $1.13^{+0.59}_{-0.47}$ | $1.12^{+0.61}_{-0.49}$ |
| $\beta_2$ | | $2.51^{+0.66}_{-0.54}$ | $2.54^{+0.65}_{-0.52}$ | $2.55^{+0.66}_{-0.53}$ | $2.53^{+0.66}_{-0.53}$ |
| $m_{\rm min}$ | $\mathrm{M}_\odot$ | $0.99^{+0.26}_{-0.32}$ | $0.99^{+0.25}_{-0.32}$ | $0.98^{+0.26}_{-0.32}$ | $1.01^{+0.24}_{-0.33}$ |
| $m_{\rm max}$ | $\mathrm{M}_\odot$ | $97.6^{+42.9}_{-17.6}$ | $95.6^{+47.9}_{-17.6}$ | $95.0^{+46.6}_{-18.0}$ | $97.5^{+43.3}_{-18.1}$ |
| $\lambda_{\rm g}$ | | $0.22^{+0.13}_{-0.09}$ | $0.22^{+0.12}_{-0.09}$ | $0.21^{+0.13}_{-0.09}$ | $0.22^{+0.13}_{-0.09}$ |
| $\lambda_{\rm g}^{\rm low}$ | | $0.69^{+0.09}_{-0.10}$ | $0.68^{+0.09}_{-0.11}$ | $0.68^{+0.09}_{-0.11}$ | $0.68^{+0.09}_{-0.10}$ |
| $\mu_{\rm g}^{\rm low}$ | $\mathrm{M}_\odot$ | $8.96^{+0.48}_{-0.73}$ | $8.83^{+0.43}_{-0.71}$ | $8.79^{+0.48}_{-0.81}$ | $8.90^{+0.52}_{-0.75}$ |
| $\sigma_{\rm g}^{\rm low}$ | $\mathrm{M}_\odot$ | $0.83^{+0.47}_{-0.26}$ | $0.80^{+0.46}_{-0.25}$ | $0.80^{+0.50}_{-0.25}$ | $0.82^{+0.46}_{-0.26}$ |
| $\mu_{\rm g}^{\rm high}$ | $\mathrm{M}_\odot$ | $25.0^{+3.6}_{-5.3}$ | $24.4^{+3.7}_{-5.7}$ | $24.2^{+3.8}_{-5.9}$ | $24.4^{+3.9}_{-5.5}$ |
| $\sigma_{\rm g}^{\rm high}$ | $\mathrm{M}_\odot$ | $9.8^{+3.4}_{-3.1}$ | $9.6^{+3.7}_{-3.2}$ | $9.6^{+3.6}_{-3.2}$ | $10.0^{+3.4}_{-3.3}$ |
| $\delta_{\rm m}^{\rm max}$ | $\mathrm{M}_\odot$ | $0.09^{+1.65}_{-0.08}$ | $0.08^{+1.64}_{-0.08}$ | $0.03^{+0.20}_{-0.02}$ | $0.10^{+1.84}_{-0.09}$ |
| $\delta_{\rm m}^{\rm min}$ | $\mathrm{M}_\odot$ | $0.08^{+0.23}_{-0.06}$ | $0.08^{+0.27}_{-0.06}$ | $0.07^{+0.22}_{-0.05}$ | $0.07^{+0.22}_{-0.05}$ |
| $A$ | | $0.45^{+0.29}_{-0.29}$ | $0.47^{+0.28}_{-0.30}$ | $0.45^{+0.29}_{-0.29}$ | $0.44^{+0.29}_{-0.29}$ |
| $m_{\rm d}^{\rm low}$ | $\mathrm{M}_\odot$ | $2.23^{+0.52}_{-0.49}$ | $2.28^{+0.49}_{-0.51}$ | $2.26^{+0.51}_{-0.51}$ | $2.25^{+0.51}_{-0.51}$ |
| $m_{\rm d}^{\rm high}$ | $\mathrm{M}_\odot$ | $7.2^{+1.2}_{-1.7}$ | $7.2^{+1.2}_{-1.7}$ | $7.3^{+1.2}_{-1.7}$ | $7.2^{+1.2}_{-1.6}$ |
| $\delta_{\rm d}^{\rm low}$ | $\mathrm{M}_\odot$ | $0.12^{+0.56}_{-0.10}$ | $0.11^{+0.56}_{-0.09}$ | $0.13^{+0.66}_{-0.10}$ | $0.13^{+0.64}_{-0.11}$ |
| $\delta_{\rm d}^{\rm high}$ | $\mathrm{M}_\odot$ | $0.13^{+0.59}_{-0.11}$ | $0.16^{+0.71}_{-0.14}$ | $0.12^{+0.68}_{-0.10}$ | $0.12^{+0.63}_{-0.10}$ |

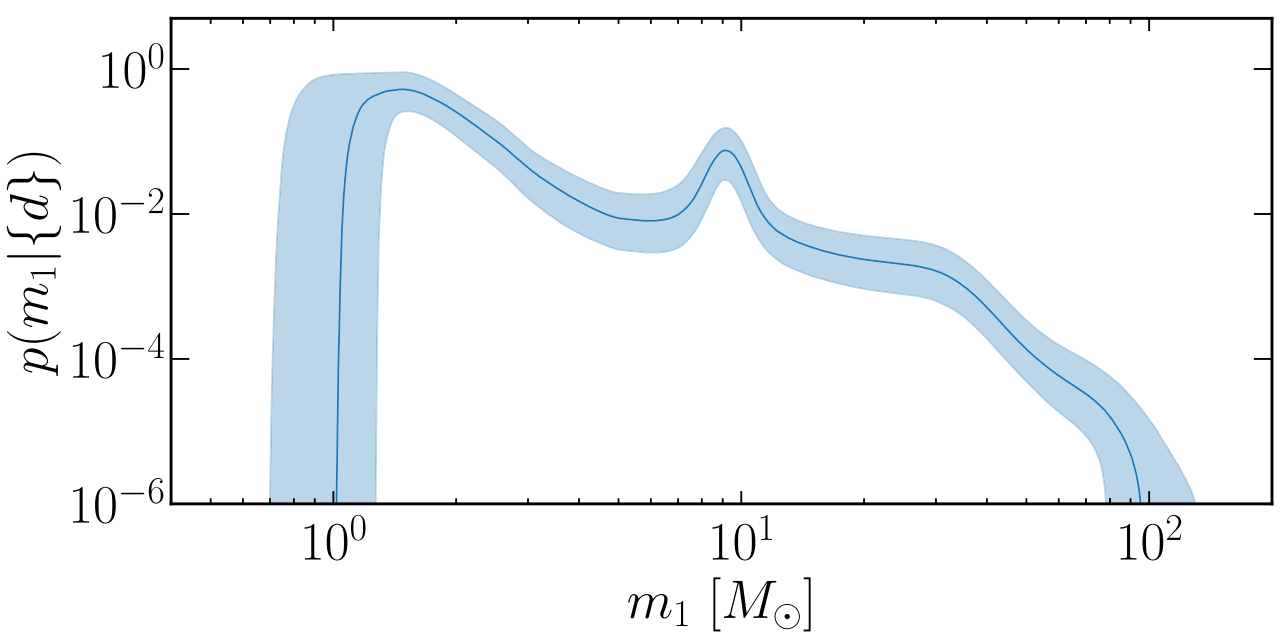

**Figure A1.** Reconstructed source-frame primary mass distribution inferred from dark sirens with median (solid line) and 1$\sigma$ (68.3 per cent) confidence interval, assuming the FULLPOP-4.0 CBC mass model.

This paper has been typeset from a TEX/LATEX file prepared by the author.